\documentclass[aps,12pt,final,notitlepage,oneside,onecolumn,
 nobibnotes,nofootinbib,superscriptaddress,noshowpacs,centertags,showkeys]
{revtex4-1}

\usepackage{amsmath}
\usepackage{amsfonts}
\usepackage[dvips]{graphicx}

\usepackage[cp1251]{inputenc}
\usepackage[T2A]{fontenc}

\makeatletter
\renewcommand \thefigure{\@arabic\c@figure}
\renewcommand \thetable{\@arabic\c@table}
\def\place@bibnumber@inl#1{#1.}%
\def\thesection       {\arabic{section}}
\def\p@section        {}
\def\thesubsection    {\thesection.\arabic{subsection}}
\def\p@subsection     {\thesection.}

\def\p@subsubsection  {\thesection\,\thesubsection\,}

\renewcommand{\date}{ }

\begin{document}

\title{
Dynamics of empty homogeneous isotropic three-dimensional spaces
}
\author{\firstname{A.~V.}~\surname{Klimenko}}
	\email{alklimenko@gmail.com}
	\affiliation{<<Business and Technology>>, Chelyabinsk, Russia}
	\author{V.~A.~Klimenko}
	\affiliation{<<Business and Technology>>, Chelyabinsk, Russia}	
	\affiliation{<<Chelyabinsk State University>>, Chelyabinsk, Russia}

\begin{abstract}
It is shown that there are seven types of solutions described in the framework of general relativity theory (GRT), the dynamics of empty homogeneous isotropic three-dimensional spaces. Solution of the equations of GRT, which describes the dynamics of a homogeneous isotropic universe, in the limiting case of vanishingly small effect of matter on the metric properties of space must go to one of them.

\end{abstract}

\maketitle
\newpage
\tableofcontents
\newpage

\section{
Introduction
}
According to general relativity theory (GRT), the geometric properties of four-dimensional spacetime described by the metric:
\begin{equation}
\label{1}
	ds^2 = g_{\mu\nu}dx^\mu dx^\nu.
\end{equation}
The metric coefficients $g_{\mu\nu}$ are functions of space-time coordinates $x_\alpha=(x_0,x_1,x_2,x_3)$, see, e.g., \cite{1,2,3,4}. The basis of general relativity is the hypothesis about the relationship of the gravitational field with the change of the metric properties of space-time. Functions $g_{\mu\nu}$ give a description of that field.

In the fundamental paper <<Principles of General Theory of Relativity>> (1916y.) \cite{5}, Einstein showed that the equations that describe the gravitational field in absence of matter can be written as:
\begin{equation}
\label{2}
	B_{\mu\nu} + \lambda\, g_{\mu\nu}B=0,
\end{equation}
where $\lambda$ --- some constant, $g^{\mu\nu}B_{\mu\nu}=B$ --- trace of the Einstein tensor $B_{\mu\nu}$, $B_{\mu\nu}$ is symmetric tensor, obtained by convolution of the $R^\rho_{\mu\sigma\tau}$ --- Riemann curvature tensor:
\begin{equation}
\label{3}
	B_{\mu\nu}=R^\sigma_{\mu\sigma\nu}.
\end{equation}
Tensor $B_{\mu\nu}$ can be written as:
\begin{equation}
\label{4}
	B_{\mu\nu}=R_{\mu\nu}- \frac 12\, R\, g_{\mu\nu},
\end{equation}
where $R_{\mu\nu}$ --- the Ricci tensor, and $R$ --- his trace, see, e.g., \cite{1,2,3,4}. The Ricci tensor has the form:
\begin{equation}
\label{5}
	R_{\mu\nu}= \frac{\partial \Gamma_{\mu\nu}^\alpha}{\partial x^\alpha} - \frac{\partial \Gamma_{\mu\alpha}^\alpha}{\partial x^\nu}+ \Gamma_{\mu\nu}^\alpha \Gamma_{\alpha\beta}^\beta - \Gamma_{\mu\alpha}^\beta \Gamma_{\nu\beta}^\alpha .
\end{equation}
The Christoffel symbols $\Gamma_{\mu\nu}^\alpha$ are defined by the formula:
\begin{equation}
\label{6}
	\Gamma_{\mu\nu}^\alpha= g^{\alpha \beta} \Gamma_{\beta , \mu\nu}= \frac 12 g^{\alpha\beta} \left( \frac {\partial g_{\beta\mu}}{\partial x^\nu} + \frac{\partial g_{\beta\nu}}{\partial x^{\mu}} - \frac {\partial g_{\mu\nu}}{\partial x^\beta}\right) .
\end{equation}

Einstein believed that the choice of the gravitational field equations in the form \eqref{2} is linked with a minimum of arbitrariness, because in addition to $B_{\mu\nu}$ has no other tensor of rank 2, which would be composed of the metric tensor $g_{\mu\nu}$ and its derivatives, which do not contain derivatives of higher order than the second, and would be linear with respect to the latter.

Einstein believed (see, \cite{5}) that the equation \eqref{2} for the gravitational field in absence of matter can be reduced to the equations:
\begin{equation}
\label{7}
	B_{\mu\nu}=0.
\end{equation}
In the general case it is not. When performing \eqref{7}, the equations \eqref{2} are automatically satisfied. At the same time, not all solutions of equation \eqref{2} are solutions of equation \eqref{7}. In the case of homogeneous isotropic empty spaces there are two possible solutions of equation \eqref{7}. The first describes a flat homogeneous three-dimensional stationary space the distance between any two points which is kept constant. The second --- curved opened a homogeneous three-dimensional space curvature radius of which increases (decreases) with speed of light.

In this paper, for example, the empty homogeneous isotropic spaces, we show that the equation \eqref{2} also have other solutions.

Empty space is seen as a limiting case of spaces filled with matter whose density is tends to zero. In this case, the solutions describing the empty homogeneous spaces are limit solutions describing the dynamics of homogeneous spaces filled with matter. In this regard, it is important to know the properties of these limiting solutions.

\section{
The initial equations}
Using the relation \eqref{4}, we find that the trace of the Einstein tensor $B=-R$, where $R$ --- trace of the Ricci tensor. Given this, from the equations \eqref{2} we find:
\begin{equation}
\label{8}
	R(1+4 \lambda)=0.
\end{equation}

Hence it follows that for all $\lambda\neq -0.25$, the scalar curvature of four-dimensional space-time $R$ is zero and the equation \eqref{2} given by the standard Einstein equations for empty space:
\begin{equation}
\label{9}
	R_\mu^\nu=0.
\end{equation}

At the same time, as you can see from \eqref{8}, with $\lambda = -0.25$ empty space may have a scalar curvature $R$ is different from zero. This means that when $\lambda = -0.25$ there may exist solutions of the equations \eqref{2} which are not solutions of the equations \eqref{9}.

Let us show that for the empty spaces the scalar curvature $R$ can not be variable. Taking the covariant derivative of the left side of the equation \eqref{2}, and noting that
\begin{equation}
	\label{v10}
\nabla_\nu \left(   R_\mu^\nu - \frac 12 R\, \delta_\mu^\nu \right)=0, 
\end{equation}
see, e.g., \cite{1, 6}, we find:

\begin{equation}
	\label{v11}
\frac{\partial  R}{\partial x^\mu}=0.
\end{equation}
This means that if $\lambda=-0.25$ scalar curvature of four-dimensional space-time $R$  although may be not equal to zero, but is a constant value. At the same time, we should note that this does not mean that the curvature of the corresponding three-dimensional space remains constant.

In the case when the scalar curvature $R$ is different from zero, using the notation:
\begin{equation}
	\label{v12}
\Lambda = - \frac 14 \,R,
\end{equation}
equation \eqref{2} can be written as:
\begin{equation}
	\label{v13}
R_\mu^\nu - \frac 12 R\, \delta_\mu^\nu = \Lambda\, \delta_\mu^\nu. 
\end{equation}

This equation is Einstein's equation with $\Lambda$-term for empty space. The constant $\Lambda$ is called the cosmological constant, see, e.g., \cite{2,7}.

Assuming that the empty spaces are homogeneous and isotropic, we find the solutions of \eqref{2}, which can describe them.

To describe the geometry of homogeneous isotropic non-stationary three-dimensional spaces is convenient to start from the geometric analogy, considering these spaces as homogeneous and isotropic three-dimensional hypersurfaces in four-dimensional space (see, e.g., \cite{1}). 

The geometry of these three-dimensional homogeneous isotropic spaces defined by the parameter $k$ and the radius of curvature $a$.

The parameter $k$ takes three values: $k=-1,\,0,\;1$. When $k=+1,\;-1,\;0$ realized cases of three-dimensional spaces with positive, negative and zero curvature, respectively. In non-stationary spaces, their radii of curvature $a$ vary over time. The metric of four-dimensional space-time can be reduced to:
\begin{equation}
\label{11}
	ds^2 = c^2 dt^2 - a^2(t) \left\lbrace  d\chi^2 + f(\chi) (d\theta^2 + \sin^2 \theta\, d \varphi^2) \right\rbrace ,
\end{equation}
where
\begin{equation}
\label{12}
f(\chi)=  
\begin{cases}
	\sin^2 \chi,&\text{at $k=+1$;}\\
	\sinh^2 \chi,&\text{at $k=-1$;}\\
	\chi^2,&\text{at $k=0$,}
\end{cases}
\end{equation}
for details see, e.g., \cite{1, 2}. 

By using the metric \eqref{11}, equations \eqref{2} in the standard way to transform to the modified cosmological Friedman equations:
\begin{equation}
\label{13}
		3\left(  \frac{\dot a ^2}{a^2}  + \frac{kc^2}{a^2}   \right) + 6 \lambda \left(   \frac{\dot a^2}{a^2}  + \frac{kc^2}{a^2}  + \frac{\ddot a}{a} \right) = 0,  
\end{equation}

\begin{equation}
\label{14}
	2 \frac{\ddot a}{a} + \frac{\dot a^2}{a^2} + \frac{kc^2}{a^2} + 6 \lambda \left(   \frac{\dot a^2}{a^2}  + \frac{kc^2}{a^2}  + \frac{\ddot a}{a}        \right) = 0.
\end{equation}

Equations \eqref{13} and \eqref{14} are consistent at the condition:
\begin{equation}
\label{15}
	\ddot a a - \dot a^2 - kc^2 =0.
\end{equation}

The formula defining the scalar curvature $R$ of four-dimensional space-time through the scale factor $a(t)$, is given by:
\begin{equation}
	\label{v20}
R=- \frac{6}{c^2 a^2} (kc^2 + a \ddot a + \dot a^2).
\end{equation}

Note that the Friedman equation \eqref{13}, \eqref{14} are invariant under transformations of the form: $a\to-a;\;\;t\to-t;\;\;t\to t+\Delta t$, where $\Delta t$ --- constant. This result is expected, because in the equation \eqref{11}, which determine the metric of the space, the scale factor $a(t)$ is included in the square, and the time $t$ is not included explicitly.

\section{
Flat empty homogeneous three-dimensional spaces
}
\subsection{
Flat three-dimensional stationary space
}
When $k=0$ and parameter $\lambda=0$, the solution of equations \eqref{13}, \eqref{14}, satisfying the condition \eqref{15}, is a function:
\begin{equation}
\label{16}
	a=a_0=const.
\end{equation}
This solution describes a stationary flat empty three-dimensional space, the distance between any two points of which is kept constant. Scalar curvature $R$ of four-dimensional spacetime in this case is zero. The condition \eqref{15} admits a solution where $a_0=0$, but the physical meaning of such a solution for us is not obvious.

\subsection{
Flat three-dimensional inflationary spaces
}
In addition to the stationary solution of \eqref{16}, when the parameter $\lambda=-0.25$, equation \eqref{13}, \eqref{14} to an empty flat three-dimensional nonstationary solutions are:
\begin{equation}
\label{17}
	a(t)=a_0\, \exp \left(  \pm \frac {t}{t_0}\right) ,
\end{equation}
where $a_0=ct_0$, $t_0$ --- a free parameter.

Solutions with the plus sign describe the exponentially expanding and contracting with a minus sign flat space. Function \eqref{17} are not solutions of the standard Einstein equations \eqref{9}. They are solutions of the Einstein equations with $\Lambda$-term \eqref{v13}. The relationship between the characteristic time $t_0$ and the cosmological constant $\Lambda$ is given by:
\begin{equation}
\label{v23}
	\Lambda = \frac{3}{c^2\,t_0^2}.
\end{equation}

According to the decisions of \eqref{17} the relative speed of convergence (expansion) of the points of three-dimensional spaces can be superluminal.

\textit{Remark.} In this paper, the fact of exponential divergence of the scale factor $a(t)$ we define by the word inflation.

\section{
Homogeneous curved empty three-dimensional spaces}
When the parameter $\lambda=-0.25$ empty three-dimensional homogeneous spaces can be not only flat, but curved. The parameter $k$, determining the type of the geometry of these spaces can be either $1$ and $-1$.

When $k=+1$ three-dimensional curved empty homogeneous isotropic space has finite volume and is closed. The curved homogeneous three-dimensional spaces of infinite volume are opened and are described by the metric \eqref{11} in which the parameter $k=-1$. Let us find the solutions of \eqref{13}, \eqref{14} corresponding to the cases $k=-1$, $\lambda=-0.25$ and $k=+1$, $\lambda=-0.25$.

\subsection{
Opened homogeneous empty three-dimensional curved spaces
}
\subsubsection{
Uniformly expanding (compressing) opened space
}
When $k=-1$, condition \eqref{15} is satisfied if the radius of curvature:
\begin{equation}
\label{21}
	a(t)=c\,t .
\end{equation}

The function \eqref{21}, describes the uniform expansion (compression) of the empty three-dimensional opened curved space at light speed. It satisfies equations \eqref{13}, \eqref{14} in the case when the scalar curvature $R=0$, and hence the value of the cosmological constant $\Lambda=0$.

In \cite{8} shows that the function \eqref{21} is the asymptotic solution, which describing well the observed properties of the Universe.

\subsubsection{
Oscillating empty three-dimensional spaces
}
In addition to solutions \eqref{21}, with $k=-1$, there are other solutions of \eqref{13}, \eqref{14} satisfying \eqref{15}. They describe the oscillating homogeneous three-dimensional curved empty opened spaces. These solutions have the form:
\begin{equation}
\label{22}
	a(t)=a_{max}\sin \frac{t}{t_1},
\end{equation}
where $a_{max}=c\,t_1$, $t_1$ --- a free parameter. There are infinitely many such solutions. They differ in amplitude $a_{max}$ and periods of oscillation $T=\pi t_1$.
 
In the case when the radius of curvature $a$ defined by \eqref{22}, the scalar curvature
 \begin{equation}
\label{27}
	R=\frac{12}{a_{max}^2} >0,
\end{equation}
and the cosmological constant $\Lambda=-3/a_{max}^2<0$. Solutions \eqref{22} are quite physical. It remains only understand the physical meaning of negative cosmological constant.

\subsubsection{
Opened inflationary spaces} 
When $k=-1$ and $\Lambda>0$ solutions describing opened empty homogeneous spaces, are as follows:
\begin{equation}
\label{28}
	a=a_2\sinh \frac{t}{t_2},
\end{equation}
where $a_2=c\,t_2$, $t_2$ --- a free parameter. The cosmological constant $\Lambda$ and the characteristic time $t_2$ are related by:
\begin{equation}
\label{29}
	\Lambda = \frac{3}{c^2\, t_2^2}.
\end{equation}

\subsubsection{
A singular point of solutions}
All three types of solutions describing homogeneous empty opened three-dimensional spaces pass through the point $a=0$. The value $a=0$ is valid in the solutions of the Friedman equations \eqref{13}, \eqref{14}. At the same time it is not clear what happens with the space when $a$, becomes zero. It is possible that at this point occurs the destruction of the space and the birth of new. We believe that in this case, it is natural to continue the solution further, suggesting that the <<new>> space repeats the evolution of the <<old>> backwards in time.

\subsection{
Closed inflationary three-dimensional homogeneous empty spaces}
When $k=+1$, equation \eqref{13}, \eqref{14} has a solution:
\begin{equation}
\label{23}
	a=a_{min}\, \cosh \frac{t}{t_3},
\end{equation} 
where $a_{min}=t_3\,c$, $t_3$ --- a free parameter. The domain of existence of solutions: $-\infty<t<+\infty$. They satisfy the initial conditions:
 \begin{equation}
\label{24}
	a(0)=a_{min}, \; \dot a (0)= 0.
\end{equation}
	
According to \eqref{23} closed homogeneous three-dimensional spaces, being born into infinity, compressed to a minimum, and then expands in a reversible manner, again go to infinity. At any moment of time $t$ the volume of the three-dimensional space is finite and is determined by the formula:
\begin{equation}
\label{25}
	V=2\,\pi^2 a^3(t).
\end{equation}
See, e.g., \cite{1}. The function \eqref{23} is the solution of \eqref{13}, \eqref{14} only when the parameter $\lambda=-0.25$. The shortcoming of solutions \eqref{23}, as in the case \eqref{17} and \eqref{28}, lies in their exponential divergence, and consequently, the complexity of the physical interpretation of these solutions. Scalar curvature of the spaces described by solutions \eqref{23}:
\begin{equation}
\label{33}
	R=-\frac{12}{a_{min}^2} <0.
\end{equation}
The corresponding value of the cosmological constant $\Lambda=3/a_{min}^2>0$.

\section{
Results}
\begin{itemize}
	\item 
It is shown that general relativity allows the possibility of the existence of seven types of empty homogeneous isotropic three-dimensional spaces. They can be not only flat but also curved, opened or closed. Character of the evolution of these spaces are schematically depicted in Figure 1.
	
\begin{figure*}[h]
\includegraphics [height=10cm]{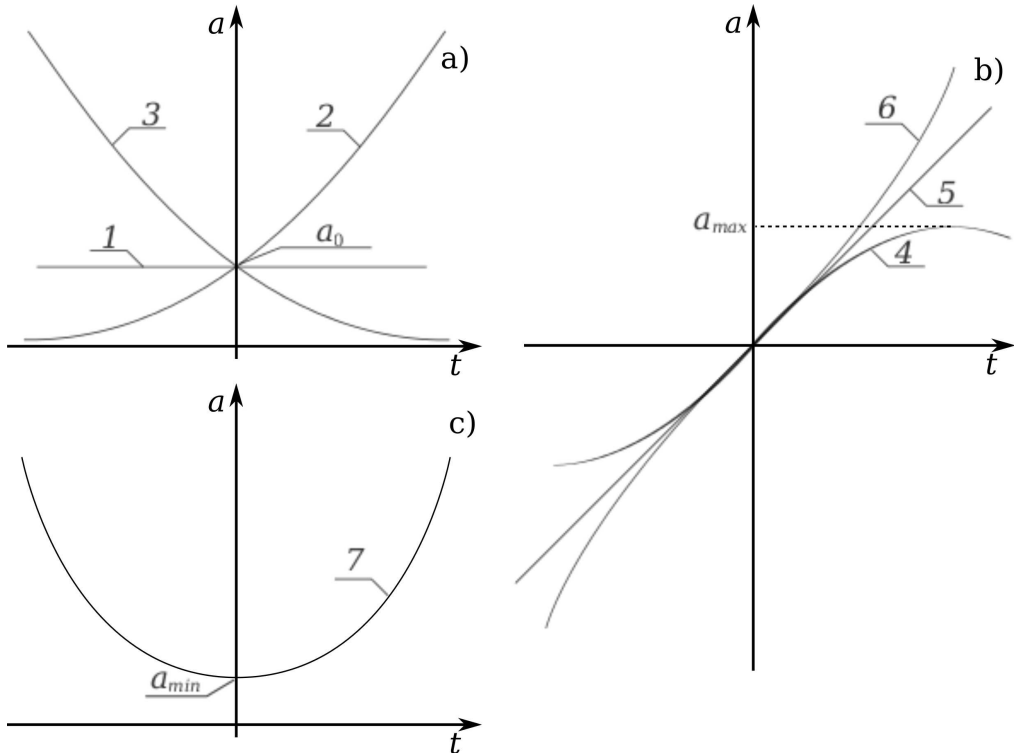} 
\caption{ 
a) flat spaces; b) curved opened spaces; c) curved closed spaces.
}
\label{pic1}
\end{figure*}	
	
\item 
The dynamics of these spaces is described by functions:
$1)\;a(t)=a_0=const;$ 
$2)\;a(t)=a_0\,\exp(t/t_0),\;t_0 =a_0/c;$ 
$3)\;a(t)=a_0\,\exp(-t/t_0);$   
$4)\;a(t)=a_{max}\sin(t/t_1),\;t_1=a_{max}/c;$ 
$5)\;a(t)=c\,t;$ 
$6)\;a(t)=a_2\sinh(t/t_2),\;t_2=a_m/c;$ 
$7)\;a(t)=a_{min}\cosh(t/t_3),\;t_3=a_{min}/c;$
	
 \item 
 Solutions of the equations of general relativity, which describe the dynamics of a homogeneous isotropic universe, in the limiting case of vanishingly small effect of matter on the metric properties of space must go to one of the above solutions describing the dynamics of empty homogeneous three-dimensional spaces.
 
 \item 
We believe that the physically interesting among them are the solutions 1, 4 and 5, which are not inflationary see Figure \ref{pic1}. We believe that proper consideration of the effect of matter on the properties of space-time can eliminate the singularity at the point $a=0$, inherent in decisions 4 and 5, see \cite{8}. Given the observed properties of the Universe, there is reason to believe that its three-dimensional space is curved and opened, and the asymptotic solution describing the cosmological expansion, is a function $a(t)=c\,t$, see \cite{8}.
 
\end{itemize}


\end{document}